\documentclass[reprint,onecolumn,aip,pof]{revtex4-1}
\usepackage{graphicx}
\usepackage{epstopdf, epsfig}
\usepackage{tikz}

\usepackage{amsmath}

\begin{document}

\title{Unsteady stretching of a glass tube with internal channel pressurisation}
\author{Gagani P. Ranathunga}
\author{Yvonne M. Stokes}
\author{Michael J. Chen}
\affiliation{School of Mathematical Sciences, The University of Adelaide, South Australia 5005, Australia}

\begin{abstract}
Mathematical modelling is used to examine the unsteady problem of heating and pulling an axisymmetric cylindrical glass tube with an over-pressure applied within the tube to form tapers with a near uniform bore and small wall thickness at the tip. To allow for the dependence of viscosity on temperature, a prescribed axially varying viscosity is assumed. Our motivation is the manufacture of emitter tips for mass spectrometry which provide a continuous fluid flow and do not become blocked. We demonstrate, for the first time, the feasibility of producing such emitters by this process and examine the influence of the process parameters, in particular the pulling force and over-pressure, on the geometry. For small values of these parameters a more uniform bore may be obtained but the geometry is highly sensitive to small fluctuations. Higher values of the parameters result in more variation in the bore size but less sensitivity to fluctuations. The best parameters depend on the accuracy of the puller used to manufacture the tapers and the permissible tolerances on the geometry. The model has wider application to the manufacture of other devices.
\end{abstract}


\maketitle

\section{Introduction}\label{sec:intro}
\begin{figure}
\centering
\tikzset{every picture/.style={line width=0.75pt}}     
       
\begin{tikzpicture}[x=0.65pt,y=0.65pt,yscale=-1,xscale=1]

\draw    (230.28,159.49) -- (453.69,158.48) ;
\draw [shift={(456.69,158.47)}, rotate = 539.74] [fill={rgb, 255:red, 0; green, 0; blue, 0 }  ][line width=0.08]  [draw opacity=0] (8.93,-4.29) -- (0,0) -- (8.93,4.29) -- cycle    ;
\draw [shift={(227.28,159.5)}, rotate = 359.74] [fill={rgb, 255:red, 0; green, 0; blue, 0 }  ][line width=0.08]  [draw opacity=0] (8.93,-4.29) -- (0,0) -- (8.93,4.29) -- cycle    ;
\draw  [draw opacity=0][fill={rgb, 255:red, 208; green, 2; blue, 27 }  ,fill opacity=1 ] (286.8,166.32) -- (389.25,166.32) -- (389.25,173.73) -- (286.8,173.73) -- cycle ;
\draw  [dash pattern={on 1.69pt off 2.76pt}][line width=1.5]  (227.92,218.42) .. controls (226.81,218.44) and (225.85,214.53) .. (225.8,209.69) .. controls (225.74,204.85) and (226.59,200.92) .. (227.71,200.9) .. controls (228.82,200.88) and (229.77,204.79) .. (229.83,209.62) .. controls (229.89,214.46) and (229.03,218.4) .. (227.92,218.42) -- cycle ;
\draw [line width=0.75]    (332.67,212.85) -- (385.05,212.13) ;
\draw [shift={(387.05,212.11)}, rotate = 539.22] [color={rgb, 255:red, 0; green, 0; blue, 0 }  ][line width=0.75]    (10.93,-4.9) .. controls (6.95,-2.3) and (3.31,-0.67) .. (0,0) .. controls (3.31,0.67) and (6.95,2.3) .. (10.93,4.9)   ;
\draw [line width=0.75]    (332.3,212.85) -- (332.3,140.85) ;
\draw [shift={(332.3,138.85)}, rotate = 450] [color={rgb, 255:red, 0; green, 0; blue, 0 }  ][line width=0.75]    (10.93,-4.9) .. controls (6.95,-2.3) and (3.31,-0.67) .. (0,0) .. controls (3.31,0.67) and (6.95,2.3) .. (10.93,4.9)   ;
\draw  [draw opacity=0][dash pattern={on 5.63pt off 4.5pt}][line width=1.5]  (230.87,183.12) .. controls (234.56,184.41) and (237.3,196.2) .. (237.11,210.5) .. controls (236.91,225.64) and (233.5,237.86) .. (229.5,237.81) .. controls (229.2,237.8) and (228.92,237.73) .. (228.64,237.6) -- (229.86,210.4) -- cycle ; \draw  [dash pattern={on 5.63pt off 4.5pt}][line width=1.5]  (230.87,183.12) .. controls (234.56,184.41) and (237.3,196.2) .. (237.11,210.5) .. controls (236.91,225.64) and (233.5,237.86) .. (229.5,237.81) .. controls (229.2,237.8) and (228.92,237.73) .. (228.64,237.6) ;
\draw  [line width=1.5]  (455.02,219.42) .. controls (453.91,219.44) and (452.96,215.54) .. (452.9,210.7) .. controls (452.84,205.86) and (453.69,201.92) .. (454.81,201.9) .. controls (455.92,201.89) and (456.87,205.79) .. (456.93,210.63) .. controls (456.99,215.47) and (456.14,219.41) .. (455.02,219.42) -- cycle ;
\draw [line width=0.75]  [dash pattern={on 4.5pt off 4.5pt}]  (225.45,201.4) -- (455.44,201.9) ;
\draw [line width=0.75]  [dash pattern={on 4.5pt off 4.5pt}]  (226.71,218.76) -- (455.65,219.42) ;
\draw [line width=1.5]    (253.8,209.92) -- (201.79,209.49) ;
\draw [shift={(197.79,209.46)}, rotate = 360.48] [fill={rgb, 255:red, 0; green, 0; blue, 0 }  ][line width=0.08]  [draw opacity=0] (11.61,-5.58) -- (0,0) -- (11.61,5.58) -- cycle    ;
\draw [line width=1.5]    (433.76,210.44) -- (486.25,212.62) ;
\draw [shift={(490.24,212.78)}, rotate = 182.37] [fill={rgb, 255:red, 0; green, 0; blue, 0 }  ][line width=0.08]  [draw opacity=0] (11.61,-5.58) -- (0,0) -- (11.61,5.58) -- cycle    ;
\draw [line width=0.75]  [dash pattern={on 4.5pt off 4.5pt}]  (456.81,247.82) -- (456.93,210.63) -- (455.4,176.36) -- (455.43,152.07) ;
\draw  [draw opacity=0][fill={rgb, 255:red, 208; green, 2; blue, 27 }  ,fill opacity=1 ] (286.17,243.06) -- (388.62,243.06) -- (388.62,250.46) -- (286.17,250.46) -- cycle ;
\draw    (287.73,257.1) -- (387.25,256.24) ;
\draw [shift={(390.25,256.21)}, rotate = 539.5] [fill={rgb, 255:red, 0; green, 0; blue, 0 }  ][line width=0.08]  [draw opacity=0] (8.93,-4.29) -- (0,0) -- (8.93,4.29) -- cycle    ;
\draw [shift={(284.73,257.12)}, rotate = 359.5] [fill={rgb, 255:red, 0; green, 0; blue, 0 }  ][line width=0.08]  [draw opacity=0] (8.93,-4.29) -- (0,0) -- (8.93,4.29) -- cycle    ;
\draw  [dash pattern={on 5.63pt off 4.5pt}][line width=1.5]  (131.92,334.58) .. controls (135.46,334.58) and (138.33,349.13) .. (138.33,367.08) .. controls (138.33,385.03) and (135.46,399.58) .. (131.92,399.58) .. controls (128.37,399.58) and (125.5,385.03) .. (125.5,367.08) .. controls (125.5,349.13) and (128.37,334.58) .. (131.92,334.58) -- cycle ;
\draw  [dash pattern={on 4.5pt off 4.5pt}] (533.6,370.99) -- (131.85,372.46) .. controls (131.2,372.47) and (130.67,370.71) .. (130.65,368.54) .. controls (130.64,366.37) and (131.16,364.61) .. (131.81,364.6) -- (533.56,363.13) .. controls (534.21,363.13) and (534.75,364.89) .. (534.76,367.06) .. controls (534.77,369.23) and (534.25,370.99) .. (533.6,370.99) .. controls (532.95,370.99) and (532.41,369.24) .. (532.4,367.07) .. controls (532.39,364.89) and (532.91,363.13) .. (533.56,363.13) ;
\draw  [dash pattern={on 4.5pt off 4.5pt}] (132.88,369.96) .. controls (131.64,369.95) and (130.64,368.33) .. (130.65,366.33) .. controls (130.66,364.34) and (131.67,362.73) .. (132.91,362.73) .. controls (134.15,362.74) and (135.15,364.36) .. (135.14,366.36) .. controls (135.13,368.35) and (134.12,369.96) .. (132.88,369.96) -- cycle ;
\draw    (282.95,336.36) -- (385.28,336.75) ;
\draw [shift={(388.28,336.76)}, rotate = 180.22] [fill={rgb, 255:red, 0; green, 0; blue, 0 }  ][line width=0.08]  [draw opacity=0] (8.93,-4.29) -- (0,0) -- (8.93,4.29) -- cycle    ;
\draw [shift={(279.95,336.35)}, rotate = 0.22] [fill={rgb, 255:red, 0; green, 0; blue, 0 }  ][line width=0.08]  [draw opacity=0] (8.93,-4.29) -- (0,0) -- (8.93,4.29) -- cycle    ;
\draw [line width=0.75]  [dash pattern={on 4.5pt off 4.5pt}]  (334.39,401.3) -- (334.21,359.13) -- (333.97,305.69) ;
\draw  [draw opacity=0][line width=1.5]  (133.32,401.49) .. controls (128.4,396.22) and (124.89,382.39) .. (124.89,366.15) .. controls (124.89,353.54) and (127.01,342.39) .. (130.26,335.56) -- (137.73,366.15) -- cycle ; \draw  [line width=1.5]  (133.32,401.49) .. controls (128.4,396.22) and (124.89,382.39) .. (124.89,366.15) .. controls (124.89,353.54) and (127.01,342.39) .. (130.26,335.56) ;
\draw  [draw opacity=0][line width=1.5]  (532.76,336.56) .. controls (538.05,342.16) and (541.83,353.75) .. (542.1,367.22) .. controls (542.39,381.66) and (538.56,394.13) .. (532.88,399.48) -- (526.13,367.62) -- cycle ; \draw  [line width=1.5]  (532.76,336.56) .. controls (538.05,342.16) and (541.83,353.75) .. (542.1,367.22) .. controls (542.39,381.66) and (538.56,394.13) .. (532.88,399.48) ;
\draw  [draw opacity=0][line width=1.5]  (530.56,399.97) .. controls (525.7,393.7) and (522.31,381.58) .. (522.11,367.6) .. controls (521.9,352.58) and (525.44,339.57) .. (530.72,333.61) -- (537.66,367.32) -- cycle ; \draw  [line width=1.5]  (530.56,399.97) .. controls (525.7,393.7) and (522.31,381.58) .. (522.11,367.6) .. controls (521.9,352.58) and (525.44,339.57) .. (530.72,333.61) ;
\draw [line width=1.5]    (504.27,366.73) -- (564.53,367.36) ;
\draw [shift={(568.53,367.4)}, rotate = 180.59] [fill={rgb, 255:red, 0; green, 0; blue, 0 }  ][line width=0.08]  [draw opacity=0] (11.61,-5.58) -- (0,0) -- (11.61,5.58) -- cycle    ;
\draw [line width=1.5]    (94.82,368.28) -- (171.39,368.59) ;
\draw [shift={(90.82,368.26)}, rotate = 0.24] [fill={rgb, 255:red, 0; green, 0; blue, 0 }  ][line width=0.08]  [draw opacity=0] (11.61,-5.58) -- (0,0) -- (11.61,5.58) -- cycle    ;
\draw [line width=0.75]  [dash pattern={on 4.5pt off 4.5pt}]  (532.9,405.55) -- (531.54,363.16) -- (531.25,328.97) -- (531.05,314.37) ;
\draw [line width=1.5]    (353.5,378.15) .. controls (237.57,375.17) and (241.06,394.74) .. (202.18,398.31) .. controls (163.29,401.87) and (146.83,399.47) .. (132.17,400.06) ;
\draw [line width=1.5]    (128.82,335.57) .. controls (244.75,332.59) and (241.26,352.17) .. (280.14,355.73) .. controls (319.03,359.29) and (335.49,356.9) .. (350.15,357.48) ;
\draw [line width=1.5]    (311.22,378.45) .. controls (427.16,375.47) and (423.66,395.04) .. (462.55,398.61) .. controls (501.43,402.17) and (517.89,399.77) .. (532.55,400.36) ;
\draw [line width=1.5]    (311.27,357.44) .. controls (427.2,360.43) and (423.71,340.85) .. (462.59,337.29) .. controls (501.48,333.72) and (517.94,336.12) .. (532.6,335.53) ;
\draw  [draw opacity=0][fill={rgb, 255:red, 208; green, 2; blue, 27 }  ,fill opacity=1 ] (283.17,344.06) -- (385.62,344.06) -- (385.62,351.46) -- (283.17,351.46) -- cycle ;
\draw  [draw opacity=0][fill={rgb, 255:red, 208; green, 2; blue, 27 }  ,fill opacity=1 ] (282.17,386.06) -- (384.62,386.06) -- (384.62,393.46) -- (282.17,393.46) -- cycle ;
\draw  [line width=1.5]  (453.31,240.27) -- (229.64,239.6) .. controls (224.88,239.59) and (221.07,226.74) .. (221.12,210.9) .. controls (221.16,195.06) and (225.05,182.23) .. (229.81,182.24) -- (453.49,182.91) .. controls (458.24,182.92) and (462.05,195.77) .. (462,211.61) .. controls (461.96,227.45) and (458.07,240.28) .. (453.31,240.27) .. controls (448.56,240.25) and (444.75,227.4) .. (444.8,211.56) .. controls (444.84,195.72) and (448.73,182.89) .. (453.49,182.91) ;
\draw  [line width=0.75]  (531.65,371.92) .. controls (530.54,371.94) and (529.61,370) .. (529.58,367.58) .. controls (529.55,365.16) and (530.43,363.18) .. (531.54,363.16) .. controls (532.66,363.15) and (533.58,365.09) .. (533.61,367.51) .. controls (533.64,369.93) and (532.76,371.91) .. (531.65,371.92) -- cycle ;
\draw  [dash pattern={on 1.69pt off 2.76pt}][line width=1.5]  (133.17,373.77) .. controls (132.06,373.79) and (131.13,371.84) .. (131.1,369.42) .. controls (131.07,367.01) and (131.95,365.03) .. (133.07,365.01) .. controls (134.18,364.99) and (135.11,366.94) .. (135.14,369.36) .. controls (135.17,371.78) and (134.29,373.75) .. (133.17,373.77) -- cycle ;

\draw (435.41,267.97) node [anchor=north west][inner sep=0.75pt]  [font=\normalsize]  {$\tilde{x} =1$};
\draw (303.51,316.06) node [anchor=north west][inner sep=0.75pt]  [font=\normalsize]  {$2H$};
\draw (300.84,399.52) node [anchor=north west][inner sep=0.75pt]  [font=\normalsize]  {$x=\tilde{x} =0$};
\draw (507.26,432.53) node [anchor=north west][inner sep=0.75pt]  [font=\normalsize]  {$\tilde{x} =1$};
\draw (506.56,406.25) node [anchor=north west][inner sep=0.75pt]  [font=\normalsize]  {$x=L( t)$};
\draw (111.12,293.9) node [anchor=north west][inner sep=0.75pt]  [font=\large]  {$t >0$};
\draw (72.38,352.06) node [anchor=north west][inner sep=0.75pt]  [font=\Large]  {$F$};
\draw (576.22,352.19) node [anchor=north west][inner sep=0.75pt]  [font=\Large]  {$F$};
\draw (310.26,258.37) node [anchor=north west][inner sep=0.75pt]  [font=\normalsize]  {$2H$};
\draw (329.35,116.1) node [anchor=north west][inner sep=0.75pt]  [font=\large,xslant=-0.02]  {$r$};
\draw (113.17,133.29) node [anchor=north west][inner sep=0.75pt]  [font=\large]  {$t=0$};
\draw (497.01,194.92) node [anchor=north west][inner sep=0.75pt]  [font=\Large]  {$F$};
\draw (178.69,192.92) node [anchor=north west][inner sep=0.75pt]  [font=\Large]  {$F$};
\draw (391.97,207.36) node [anchor=north west][inner sep=0.75pt]  [font=\large]  {$x$};
\draw (432.72,245.57) node [anchor=north west][inner sep=0.75pt]  [font=\normalsize]  {$x=L_{0}$};
\draw (304.12,220.92) node [anchor=north west][inner sep=0.75pt]  [font=\normalsize]  {$x=\tilde{x} =0$};
\draw (300.22,138.41) node [anchor=north west][inner sep=0.75pt]  [font=\normalsize]  {$2L_{0}$};

\end{tikzpicture}

\caption{{The tapering process}}
    \label{fig:1}
\end{figure}
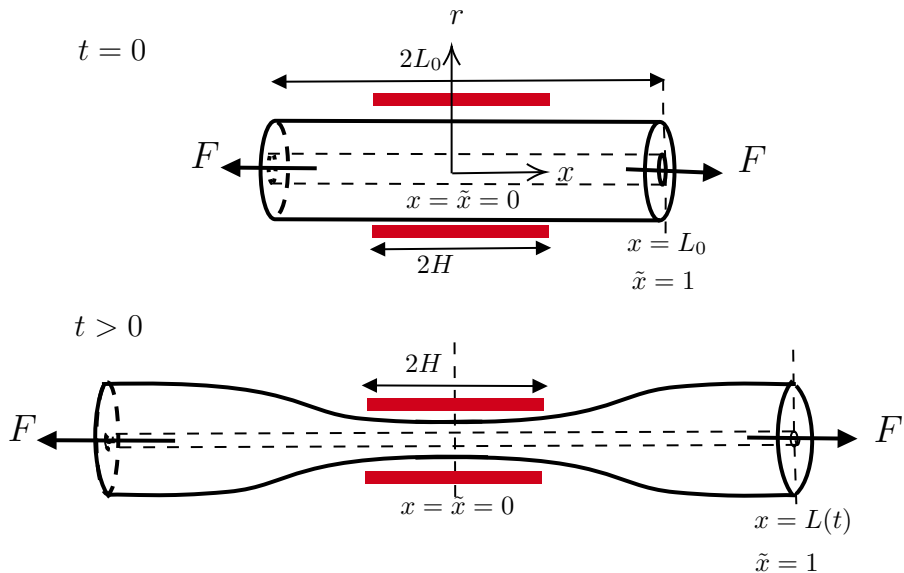

We consider the stretching of an axisymmetric glass micro-capillary with initial external radius $R_0$, internal radius $\phi_0 R_0$, $0<\phi_0<1$, and length $2L_0$, by heating over a length $2H$ in the centre and pulling both ends in opposite directions with a force of prescribed magnitude $F$; see figure~\ref{fig:1}. The initial geometry is, typically, slender, i.e. $R_0/L_0\ll 1$. An over-pressure $P$ above atmospheric pressure may be applied within the internal channel. This is an unsteady extensional-flow problem. As the glass temperature increases the viscosity reduces until the central heated region extends and thins rapidly to yield an hour-glass shape. During stretching the cross-sectional geometry will also deform under the effects of surface tension and applied pressure, with pressure counteracting the closure of the channel by surface tension and, perhaps, further expanding it.  When cooled and cut transversely at the centre, two identical tapered capillaries are obtained. 

Our focus here is on the manufacture of tapered capillaries by this unsteady heat and pull method for use as emitter tips in mass spectrometers used to analyse liquid samples \citep{valaskovic1995attomole,wilm1996analytical,gibson2009nanoelectrospray}. To improve the sensitivity of biological and chemical mass spectrometry and avoid clogging of the tip, a small near-uniform bore of $10-20\,\mu$m is desirable with the external wall tapering over a length of around 5\,mm from $75-150\,\mu$m in radius to a sharp end with radius around $8-15\,\mu$m. Over-pressure in the channel is essential to obtain a near-uniform bore.  
In this paper, we derive and use a mathematical model to investigate the interplay between viscosity (effectively temperature), pulling force, surface tension and over-pressure on the geometry of the final emitter tip.
We note here that our model might be modified for application to the manufacture of components for whispering gallery resonator sensing, for example, thin-walled hollow microbottles \citep{bianucci2016optical}. However, this is for future investigation.

A problem which strongly resembles that considered here is the manufacture of glass micro-electrodes from axisymmetric tubing by a heat and pull process essentially identical to that just described, excepting that no over-pressure is applied in the channel. This was studied by \citet{huang2003heat,huang2007formation}  
for both fixed and variable pulling forces using a coupled flow and temperature model which neglected surface tension  and, as already mentioned, did not include channel over-pressure. It is just one application of a class of extensional flow problems that have received significant attention, particularly in the last 30 decades, which concern the stretching of a slender cylinder to form a long thin fibre/thread. Other applications include the spinning of textile threads, rheological measurement, and the manufacture of optical fibres. For these problems, model derivation using an asymptotic approach which exploits the slenderness of the initial geometry has been used widely. This work is used in our modelling of emitter tip manufacture and, hence, is reviewed below.

\citet{matovich1969spinning} were the first to derive and use a one dimensional model to investigate the quasi-steady spinning of axisymmetric textile threads having no internal structure. More recent similar models include \citet{wylie2007extensional} and \citet{wylie2007thermal}.  The quasi-steady drawing of tubes, with an over-pressure inside the tube but neglecting surface tension, was first investigated by \citet{pearson1970}, again using a one dimensional model, in studying the blowing of tubular films. Motivated by the steady-state drawing of microstructured optical fibres in a draw tower, \citet{fitt2001modeling, fitt2002mathematical} developed an asymptotic model of axisymmetric capillary drawing and undertook a preliminary examination of the competition between surface tension and over-pressure in the tube. Asymptotic studies of unsteady stretching of axisymmetric viscous threads by pulling of the ends or under gravity include \citet{dewynne1989mathematical},
 \citet{stokes2000,StokesHajekTuck2011}, \citet{stokes2004role}, \citet{HajekStokesTuck2007}, \citet{wylie2007extensional}, \citet{wylie2007thermal,wylie2011stretching,wylie2016evolution}, \citet{howell2007stretching} and \citet{wylie2015asymptotic}. All of this work assumed negligible surface tension and a viscosity that is constant or varies axially in a prescribed manner. \citet{he2016extension} developed a coupled flow and temperature model with non-negligible surface tension.

A formal derivation of the asymptotic fluid-flow equations for unsteady drawing of nonaxisymmetric fibres was first given by \citet{dewynne1992systematic} 
assuming negligible inertia and surface tension and a viscosity that is at most a function of time or axial position. This work showed the simplification of the solution process brought by moving from the laboratory Eulerian reference frame to a Lagrangian reference frame, and formally demonstrated that, in the absence of surface tension, the shape of the cross-section will be preserved though it may change in scale and rotate. This model was later modified to include inertia, with inclusion of surface tension and its role in changing the cross-sectional shape  briefly discussed \citep{dewynne1994slender}. 

Building on this work and in the context of steady fibre drawing, \citet{cummings1999evolution}  developed a leading-order model for solid non-axisymmetric geometries with non-negligible surface tension and constant viscosity. This is comprised of a one-dimensional ordinary differential equation problem for axial stretching along with a two-dimensional Stokes flow problem for the flow and deformation in the cross section. Arguably the most significant aspect of this work was the introduction, in the Lagrangian reference frame, of a transformation from physical time to so-called `reduced time' to modify the kinematic condition on free boundaries in the cross-plane problem so as to render it  a classical two-dimensional Stokes flow problem driven by unit surface tension, which is readily solved.  \citet{griffiths2007surface,griffiths2008mathematical} applied this model to steady drawing of non-axisymmetric thin-walled tubes subject to surface tension but with no pressurisation of the internal channel. They also added temperature to the model, assuming temperature to be uniform in the cross section. \citet{stokes2014drawing} extended the model to fibres of arbitrary geometry; a key finding of this last work is the ability to express the axial stretching problem in terms of the reduced time variable of the cross-plane problem so that both axial and cross-plane problems can be solved independently to describe the final fibre geometry. They also showed that, for steady fibre drawing where only the geometry of the initial preform and final fibre are important, not the geometry through the entire neck-down region from preform to fibre, measurement of pulling tension obviates the need to model temperature. To this model \cite{chen2015microstructured} added active pressurisation of internal channels yielding a model with non-negligible surface tension and pressure for fibres of arbitrary geometry in which the axial-stretching and cross-plane problems are fully coupled. 

For a complete model of steady or unsteady fibre drawing either the viscosity must be known or a temperature model must be coupled to the flow model and a temperature-viscosity relation known. \citet{fitt2002mathematical} and \citet{griffiths2008mathematical} derived temperature models assuming the leading-order temperature to be uniform in a cross-section. A complete formal derivation of the temperature model, demonstrating leading-order uniformity in the cross-section, is given by \citet{he2016extension} for unsteady stretching of axisymmetric threads and by \cite{stokes2019coupled} for steady drawing of fibres with arbitrary geometry.

While the competition between surface tension and over-pressure has been considered in some detail for steady fibre drawing \citep{chen2015microstructured}, this remains to be done for unsteady fibre drawing and it is this problem that is addressed in this paper in order to determine whether emitter tips for mass spectrometry with (near) uniform bore can be manufactured by the heat and pull method. For simplicity we will here assume that the viscosity is given by a known function of axial position only and leave coupling of the flow model with a temperature model to a future publication. As will be seen, the model is similar to that of \cite{chen2015microstructured} for steady fibre drawing but the unsteady nature of the manufacturing process requires a very different the solution method. 

The remainder of this paper is structured as follows. In section~\ref{sec:math_model} we outline the key steps in the derivation of the unsteady model and, in section~\ref{sec:numerical_procedure}, we describe the numerical solution method. In section~\ref{sec:solutions} we give and discuss numerical solutions for a realistic prescribed  axially varying viscosity function, and then explore the parameter space suitable for the manufacture of emitter tips with near-uniform bore. We conclude the paper  in section~\ref{sec:conclusions} with a summary of our key findings and a discussion of future improvements. 

\section{Mathematical model}\label{sec:math_model}

As shown in figure~\ref{fig:1}, we use a polar coordinate system in which $x$ is the axial coordinate, $r$ is the radial coordinate and the origin is at the centre of mass of the tube. At time $t$ the tube has length $2L(t)$ while the external radius and aspect ratio are $R(x,t)$ and $\phi(x,t)$, respectively. The cross sectional area of the capillary tube is denoted by $S(x,t)=\pi R^2(1-\phi^2)$. Initially, the tube has uniform cross-sectional area, external radius and aspect ratio, denoted by $S(x,0)=S_0$, $R(x,0)=R_0$ and $\phi(x,0)=\phi_0$.  The tube is heated at its centre over a fixed length $2H$ and pulled at both ends with constant force $F$. A constant pressure $P$ may be applied within the inner channel to prevent its closing. Symmetry enables our consideration of just half of the tube, $0\le x\le L(t)$, equivalent to an axisymmetric tube held fixed at $x=0$, heated over $0\le x\le H$, and pulled with force $F$ at the end $x=L(t)$. We denote the fluid velocity by $\mathbf{u}=(u,v)$, where $u(\mathbf{x},t)$ and $v(\mathbf{x},t)$ are the axial and radial components of the velocity at position $\mathbf{x}=(x,r)$ and time $t\ge 0$, while $p(\mathbf{x},t)$ denotes the pressure in the fluid. The softened glass is assumed to be an incompressible Newtonian fluid with  constant density $\rho$ and surface tension $\gamma$; however, the viscosity $\mu$ of glass depends strongly on temperature and so is, in general, a function of position and time. The glass tube we consider for this study has an internal channel with radius around 3.6\,$\mu$m and an external radius around 60\,$\mu$m. Typically the tapering process should yield emitter tips which taper over a length of around 5mm such that the wall thickness at the very tip is around 6\,--\,8\,$\mu$m.

\subsection{Nondimensionalisation}

For convenience, we first define $\chi(x,t)=\sqrt{S(x,t)}$ and $\chi_0=\sqrt{S_0}$. As is justified by the parameters in table~\ref{table:1}, we assume the capillary to be slender such that $\epsilon=\chi_0/H \ll 1$. This enables a  long-wavelength approximation as has been previously used in the modelling of fibre drawing \citep{dewynne1994slender,cummings1999evolution,fitt2002mathematical,griffiths2008mathematical,stokes2014drawing,chen2015microstructured,tronnolone2016gravitational,stokes2019coupled}. We define scaled variables,  denoted by  primes, as follows:
\begin{equation}
(x,r)=H(x',\epsilon r' ),  \quad (u,v)=\frac{\gamma H}{\mu_{hot}\chi_0}( u', \epsilon v' ), \quad  t=\frac{\mu_{hot} \,  \chi_0}{\gamma}t', \quad p=\frac{\gamma}{\chi_0}p', 
\end{equation}
along with the scaled force and pressure parameters
\begin{equation}
F'=\frac{F}{6 \gamma \chi_0}, \quad P'=\frac{\chi_0}{\gamma}P. \label{scalingF_P}
\end{equation}
Note that, using the physical parameters of table~\ref{table:1}, the axial velocity scale $\gamma H/(\mu_{hot}\chi)\sim 0.1-1\,$ mm/s which is reasonable for the pulling of emitter tips.

\begin{table}
  \begin{center}
\def~{\hphantom{0}}
  \begin{tabular}{lccc}
      Description  & Symbol   &  Value & SI unit \\[3pt]
      Initial length   & $L_0$ & $\sim 5\times 10^{-3}$ & m\\
      Surface tension & $\gamma$ & $0.3$ & Nm$^{-1}$\\ 
       Minimum viscosity & $\mu_{hot}$ & $10^{4}-10^5$ & Pa\,s \\
        Density & $\rho$ & 2200 & kgm$^{-3}$ \\
        Initial external radius  & $R_0$ & $6\times 10^{-5}$ & m \\ 
        Initial aspect ratio  & $\phi_0$ & 0.06 & - \\ 
        Initial cross sectional area & $S_0$ & $1 \times 10^{-8} $ & m$^2$ \\
        Heated length & $H$ & $\sim 3\times 10^{-3}$ & m\\
        Pressure & $P$ & To be determined & Pa\\
        Force & $F$ & To be determined & N\\
  \end{tabular}
  \caption{Physical parameters relevant to manufacture of emitter tips}
  \label{table:1}
  \end{center}
\end{table}

As already mentioned, the viscosity is strongly dependent on temperature and, hence on position and time. However, as shown by \citet{he2016extension} and \citet{stokes2019coupled}, for a slender capillary the temperature is (to leading order) uniform in a cross-section at a point in time so that we consider the viscosity to be independent of the radial coordinate $r$ and, hence, a function of $x$ and $t$. Then, defining $\mu_{hot}$ to be the viscosity of the glass when it is hottest, we define the scaled viscosity as
\begin{equation}
\mu'= \mu(x,t)/\mu_{hot}. 
\end{equation}
The Reynolds number for our problem is very small such that inertial terms may be neglected, being
\begin{equation}
\Re=\frac{\rho\gamma H^2}{\mu_{hot}^2\chi_0}\sim 10^{-7}\ll 1.
\end{equation}

\subsection{Axi-symmetric model equations}

A detailed description of the asymptotic model derivation method we adopt is given for steady fibre drawing in \citet{cummings1999evolution} and \citet{stokes2019coupled}, and for unsteady fibre drawing in \citet{dewynne1994slender}.
We allow that the internal channel may be pressurised. Internal channel pressure is included as described by \citet{chen2015microstructured} for steady fibre drawing. In view of the detailed description given in the literature already referenced, we here simply give our model. Note that from here on all variables and parameters are assumed to be dimensionless quantities and we drop the primes.

First we label each cross-section of the fibre with its initial position $\tilde x$ defined by
\begin{equation}
\tilde{x} = x (\tilde{x}, 0),
\end{equation}
where $x(\tilde x,t)$ is the position of cross-section $\tilde x$ at time $t$. 
The relationship between $x$ and $\tilde x$ is defined by
\begin{equation}
\frac{\partial x}{\partial \tilde x} = \frac{\chi^2(\tilde x, 0) }{\chi^2(\tilde x, t)},\quad x(0,t)=0. \label{x-transformation}
\end{equation}
This allows for a cylinder with initial arbitrary geometry but, since our initial geometry is a tube of uniform cross-sectional area, we set $\chi^2(\tilde x,0)=1$.

We also introduce the ``reduced time'' variable $\tau(\tilde x,t)$ \citep{stokes2014drawing},
\begin{equation}
 \frac{\partial\tau}{\partial t} = \frac{1}{\mu(\tilde x,t)\, \chi(\tilde{x},t)},\quad \tau(\tilde x,0)=0,  \label{time-tau}
 \end{equation}
where we have set the dimensionless surface tension parameter to unity in line with the scaling of our problem.

With these transformations, the dimensionless model equations for the cross-sectional area $S(\tilde x,\tau)=\chi^2(\tilde x,\tau)$ and aspect ratio $\phi(\tilde x,\tau)$ of a slender axisymmetric capillary held fixed at $\tilde x=0$, pulled with force $F$ at $\tilde x=L_0/H$, and with pressure $P$ in the inner channel are
 \begin{align}
\frac{\partial\chi}{\partial\tau} &= \frac{\sqrt{\pi}}{6} \chi \sqrt{\frac{1+\phi}{1-\phi}} - {F},\quad \chi(\tilde x,0)=1, \label{stretching}\\
 \frac{\partial\phi}{\partial\tau}& = -\frac{\sqrt{\pi}}{2} (1+ \phi)^{3/2} (1-\phi)^{1/2} + \frac{P}{2}\phi \chi,\quad \phi(\tilde x,0)=\phi_0, \label{cross-section-deformation}
\end{align}
along with \eqref{x-transformation} and \eqref{time-tau}.
The outer radius of the tube is, then, given by
\begin{align}
R(\tilde{x},\tau)=\frac{\chi}{\sqrt{\pi ( \, 1 - \phi^{2})}}
\end{align} 
while the inner radius is $\phi R$.
Observe that, because initially the fibre has uniform cross-sectional area and aspect ratio,  these equations have no explicit $\tilde x$-dependence and may be solved independently of $\tilde x$, to obtain the evolution of the geometry with $\tau$ at every $\tilde x$. 

Equations \eqref{stretching} and \eqref{cross-section-deformation} may be solved first, independently of \eqref{time-tau} and \eqref{x-transformation}, for $\chi(\tilde x,\tau)$ and $\phi(\tilde x,\tau)$. Equation \eqref{stretching} fundamentally describes the change in cross-sectional area with $\tau$ due to stretching, whereas \eqref{cross-section-deformation} describes the evolution of the cross-sectional shape (i.e the aspect ratio) with $\tau$ due to surface tension and pressure. For $P\ne 0$, these two equations are fully coupled and, so, need to be solved simultaneously. For $P=0$ \eqref{cross-section-deformation} decouples from \eqref{stretching} and the solution may be found analytically as in \cite{stokes2014drawing} and \cite{tronnolone2016gravitational}. Specifically, defining  

\begin{equation}
\alpha = \sqrt{\frac{1-\phi}{\pi (1+\phi)}},
\end{equation}
the equations (with $P=0$) may be written
\begin{align}
\frac{\partial\chi}{\partial\tau}-\frac{\chi}{6\alpha}=-F\quad
\quad
\frac{\partial\alpha}{\partial\tau}=\frac{1}{2}, \label{zero_pressure_geometry}
\end{align}
and the solution is readily found to be
\begin{align}
\alpha&=\alpha_0\left(1+\frac{\tau}{2\alpha_0}\right),\label{alpha-tau} \\
\chi&=\left(\frac{\alpha}{\alpha_0}\right)^{1/3}\left\{1-3F\alpha_0\left[\left(\frac{\alpha}{\alpha_0}\right)^{2/3}-1\right]\right\},\label{chi-P0}\\
\phi& = \frac{1-\pi\alpha^2}{1+\pi\alpha^2}.
\end{align} 

A number of events are possible during the pulling process, depending on the problem parameters, namely closure of the internal channel when $\phi\rightarrow 0$, bursting of the tube when $\phi\rightarrow 1$, and breaking of the fibre defined as $\chi\rightarrow 0$. In the case of zero pressure for which $\phi$ necessarily decreases with $\tau$, bursting cannot occur and exact expressions may be found for the reduced times at which hole closure and fibre breaking occur. More generally, these events must be identified as part of the solution process. None of them are permissible in pulling of emitter tips and, though solution may be continued beyond hole closure, solution will be stopped should any one of these events occur.

To determine the physical geometry of the tube at physical position $x$ and physical time $t$ we must solve \eqref{x-transformation} and \eqref{time-tau} for $t(\tilde x,\tau)$ and $x(\tilde x,\tau)$. This is not at all straightforward because the viscosity $\mu$ is assumed to be a known function of physical position $x$ and time $t$, that is the dependent variables for which we wish to solve, rather than of the independent variables of our model, $\tilde x$ and $\tau$.  Writing \eqref{time-tau}  and \eqref{x-transformation} in the form
\begin{align}
\int_0^t\frac{1}{\mu(x,\eta)}d\eta = \int_0^\tau\chi(\tilde{x},\tau)d\tau,\label{time-tau-integral}\\
x(\tilde x,t)=\int_0^{\tilde x}\frac{1}{\chi^2(\xi,t)}d\xi,\label{x-tildex-integrated}
\end{align}
it is apparent that, although we can evaluate the right-hand-side of \eqref{time-tau-integral}, we need $x(\tilde x,\eta),\,0\le \eta\le t$ to determine $t(\tilde x,\tau)$ for each cross-section $\tilde x$ using \eqref{time-tau-integral}, while \eqref{x-tildex-integrated} shows that we need $\chi(\xi,t),\,0\le\xi\le\tilde x$ at fixed time $t$ to find $x(\tilde x,t)$. Thus these equations are fully coupled in a complex manner, requiring a solution method different from those
used to solve previously considered steady and unsteady fibre drawing problems.

\section{Full numerical solution procedure}\label{sec:numerical_procedure}
We describe our solution method assuming no closure of the internal channel, bursting of the tube, or breaking of the fibre, and for $P>0$ so that all equations require numerical solution. 

We have two coordinate systems with which to deal, the original (dimensionless) physical coordinates $(x,t)$ and the transformations of these $(\tilde x,\tau)$.
Recalling that \eqref{stretching} and \eqref{cross-section-deformation} are independent of $\tilde x$, we discretize the $\tau$-space by defining
\begin{align}
\tau_i & = i\Delta\tau,\quad i=0,1,\ldots, I,
\end{align}
where $\Delta\tau$ is the reduced-time step, and solve these equations using the {\em Matlab} ODE solver `ode45' to obtain 
\begin{equation}
\chi(\tilde x,\tau_i)=\hat\chi_i,\quad \phi(\tilde x,\tau_i)=\hat\phi_i.\label{initial-xtilde-tau-solution}
\end{equation}
`Events' are defined for the {\em Matlab} solver to stop solution should the internal channel close, the tube burst, or the fibre break. Thus, $\tau_I\le\tau_{\max}$ where $\tau_{\max}$ is the maximum possible value of $\tau$ at which $\phi=0$, or $\phi=1$, or $\chi=0$.

As seen from \eqref{time-tau-integral}, the time $t(\tilde x,\tau)$ corresponding to a fixed value $\tau>0$ will differ between cross-sections that have seen a different viscosity history. Because of this and the fact that the physical spatial domain is increasing with time, it is most convenient to find our solution in terms of $(\tilde x,t)$ coordinates. Thus we choose a spatial grid spacing $\Delta\tilde x$ and a time step $\Delta t$ and  define the discrete coordinates 
\begin{align}
\tilde x_j &= j\Delta\tilde x,\quad J=0,1,\ldots, J,\\
t_k &= k\Delta t,\quad k=0,1,\ldots, K,
\end{align}
where $\tilde x_J=L_0/H$ and $t_K$ is the final time to which we wish to compute.

We now use a time-stepping procedure to obtain the solution $\chi_{j,k}=\chi(\tilde x_j,t_k)$, $\phi_{j,k}=\phi(\tilde x_j,t_k)$, $\tau_{j,k}=\tau(\tilde x_j,t_k)$, $x_{j,k}=x(\tilde x_j,t_k)$, and $\mu_{j,k}=\mu(\tilde x_j,t_k)$ for $k=1,2,\ldots,K$, where
 $\chi_{j,0}=1$, $\phi_{j,0}=\phi_0$, $\tau_{j,0}=0$, $x_{j,0}=\tilde x_j$, and $\mu_{j,0}=\mu(\tilde x_j,0)$. Assuming the solution to be known at time $t_k$, our algorithm for obtaining the solution at time $t_{k+1}$ is as follows.
\begin{enumerate}
\item Forward differencing of \eqref{time-tau} yields
\begin{equation}
\tau_{j,k+1}=\tau_{j,k}+\frac{\Delta t}{\mu_{j,k}\chi_{j,k}},\quad j=0,1,\ldots,J.
\end{equation}

\item Determine $\chi_{j,k+1}$ and $\phi_{j,k+1}$ for $j=0,1,\ldots,J$ by interpolation of the previously obtained solution $(\tau_i,\hat\chi_i,\hat\phi_i)$ to \eqref{stretching} and \eqref{cross-section-deformation}. {\em Matlab's} `deval' function was used for this.
\label{interp-step}

\item Using the trapezoidal rule to integrate \eqref{x-tildex-integrated}, or an equivalent discretization of \eqref{x-transformation}, yields
\begin{equation}
x_{j+1,k+1}=x_{j,k+1}+\frac{\Delta\tilde x}{2}\left(\frac{1}{\chi^2_{j+1,k+1}}+\frac{1}{\chi^2_{j,k+1}}\right),\quad j=0,1,\ldots,J-1,
\end{equation}
where $x_{0,k+1}=0$.

\item Lastly, the viscosity is updated, i.e.
\begin{equation}
\mu_{j,k+1}=\mu(x_{j,k+1},t_{k+1}).
\end{equation}
\end{enumerate}
We now increment the time step and repeat this procedure through to time $t_K$. It is clear from the algorithm that we must obtain the solution \eqref{initial-xtilde-tau-solution} over a sufficiently large $\tau$-range, i.e. for a sufficiently large value $I$, to enable the interpolations of Step~\ref{interp-step} of the above time-stepping procedure for all $\tau_{j,k}$. However, should some $\tau_{j,k}$ exceed the maximum possible value such that the internal channel closes, the tube bursts, or the fibre breaks, we simply stop our solution procedure and adjust the final time $t_K$.

In this way we are able to determine the geometry of an emitter tip at the final time $t_K$ for given force $F$, pressure $P$ and viscosity $\mu(x,t)$. We use $\Delta\tau= \Delta\tilde x=10^{-3}$ and $\Delta t = 10^{-4}$ to give results that are sufficiently accurate in a reasonable time. 

Although our primary interest in this paper is the pulling of tapered emitter tips with constant bore, which requires channel pressurisation, the analytical solution for $\phi$ and $\chi$ at reduced time $\tau$ in the case of zero pressure, discussed in section~\ref{sec:math_model} above, was used to validate the numerical determination of $\chi_{j,k+1}$ and $\phi_{j,k+1}$ at time $\tau_{j,k+1}$ by direct computation of these values. The agreement between the analytical and numerical methods was excellent. 

To illustrate the numerical procedure we use the simple
viscosity profile   
\begin{equation}
\mu(x)=\left\{\begin{array}{ll}
1, & 0\le x<1\\
100, & x\ge 1,
\end{array}\right.\label{viscosity_profile}
\end{equation}
 i.e. the viscosity is independent of time and piece-wise constant in space 
 with a step change at $x=1$ from $\mu=1$ (hot) in the heated region to $\mu=100$ (cold) in the cold region beyond. We consider the case of  positive pressure in the fibre channel, as is required for maintenance of the internal channel. This fully couples \eqref{stretching} and  \eqref{cross-section-deformation}, requiring that we use the full numerical scheme. 

\begin{figure}
     \centering
    \includegraphics[width=0.80\textwidth]{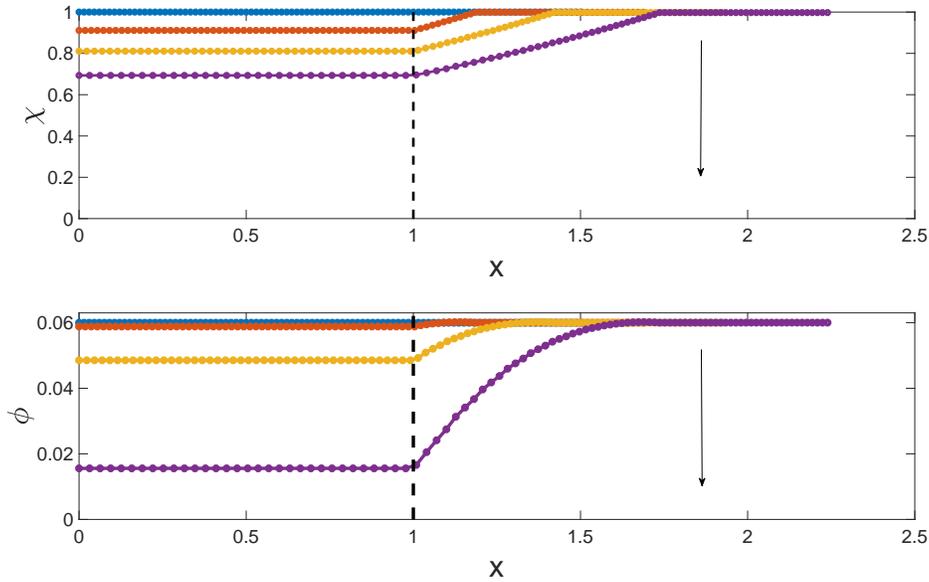}
    \caption{The evolution over time of the geometry of an axisymmetric fibre with initial (dimensionless) geometry given by $L(0)=1.5$, $\chi(x,0)=1$ and $\phi(x,0)=0.06$. The square root of the cross-sectional area $\chi(x,t)$ (top) and aspect ratio $\phi(x,t)$ (bottom) are shown against 
     axial position $x$ at times $t=0,0.05,0.1,0.15$, where the arrow shows the direction of increasing $t$. The filled dots denote 101 individual cross-sections, equispaced at $t=0$ and labeled by $\tilde x$.  Pulling with force $F=2$ and pressure $P=32$ results in a decrease in both $\chi$ and $\phi$ for all cross-sections that spend time in the heated region $0\le x<1$, the extent of which is indicated by the black dashed line. Beyond $x=1$ cross-sections are solid (the viscosity is large) so that no further deformation can occur. 
     }   
     \label{fig:2}
    \end{figure}
Figure~\ref{fig:2} shows the change in the square root of the cross sectional area $\chi(x,t)$ and aspect ratio $\phi(x,t)$ with axial position $x$ at different times $t$ where we have chosen $F=O(1)$ ( i.e. of a similar magnitude to surface tension) while $P$ is an order of magnitude larger.  The filled dots correspond to individual cross-sections and are used to visualise their motion and aid understanding of the evolution of the geometry.  Initially, all these cross-sections have unit area $\chi^2(x,0)=1$ and are spaced uniformly along the length of the tube. For the chosen parameters it is evident that inside the heater where the viscosity is sufficiently small ($\mu=1$) the cross-sections reduce in area and aspect ratio, and are also shifted in the direction of pulling as the length of the capillary increases. Because of the spatial uniformity of the viscosity in the heated region, all cross-sections that remain in this region throughout the draw undergo the same deformation, resulting in a tubular tip.  When a cross-section leaves the heater its viscosity becomes large ($\mu=100$) and deformation ceases so that its geometry remains constant from this point in time. However the cross-section continues to move in the direction of pulling due to the stretching in the heated region. All cross-sections that were initially outside the heated region undergo no deformation and retain unit area but are shifted in the direction of pulling. The cross-sections that are heated for only some part of the draw time result in a taper from the undeformed part of the tube to the tubular tip. The final geometry is shown in figure~\ref{fig:4} (left). Noting that for zero surface tension and pressure the aspect ratio $\phi$ of a cross-section will not change over time \citep{dewynne1994slender,cummings1999evolution}, the decrease in the aspect ratio seen here shows that the pressure $P$ is not sufficient to overcome the effect of surface tension, let alone increase the aspect ratio to maintain the original bore size as desired for emitter tips for mass spectrometry. 

\section{Pulling emitter tips}\label{sec:solutions}
\subsection{Physically realistic viscosity}

 \begin{figure}
 \centering
    \includegraphics[width=0.480\textwidth]{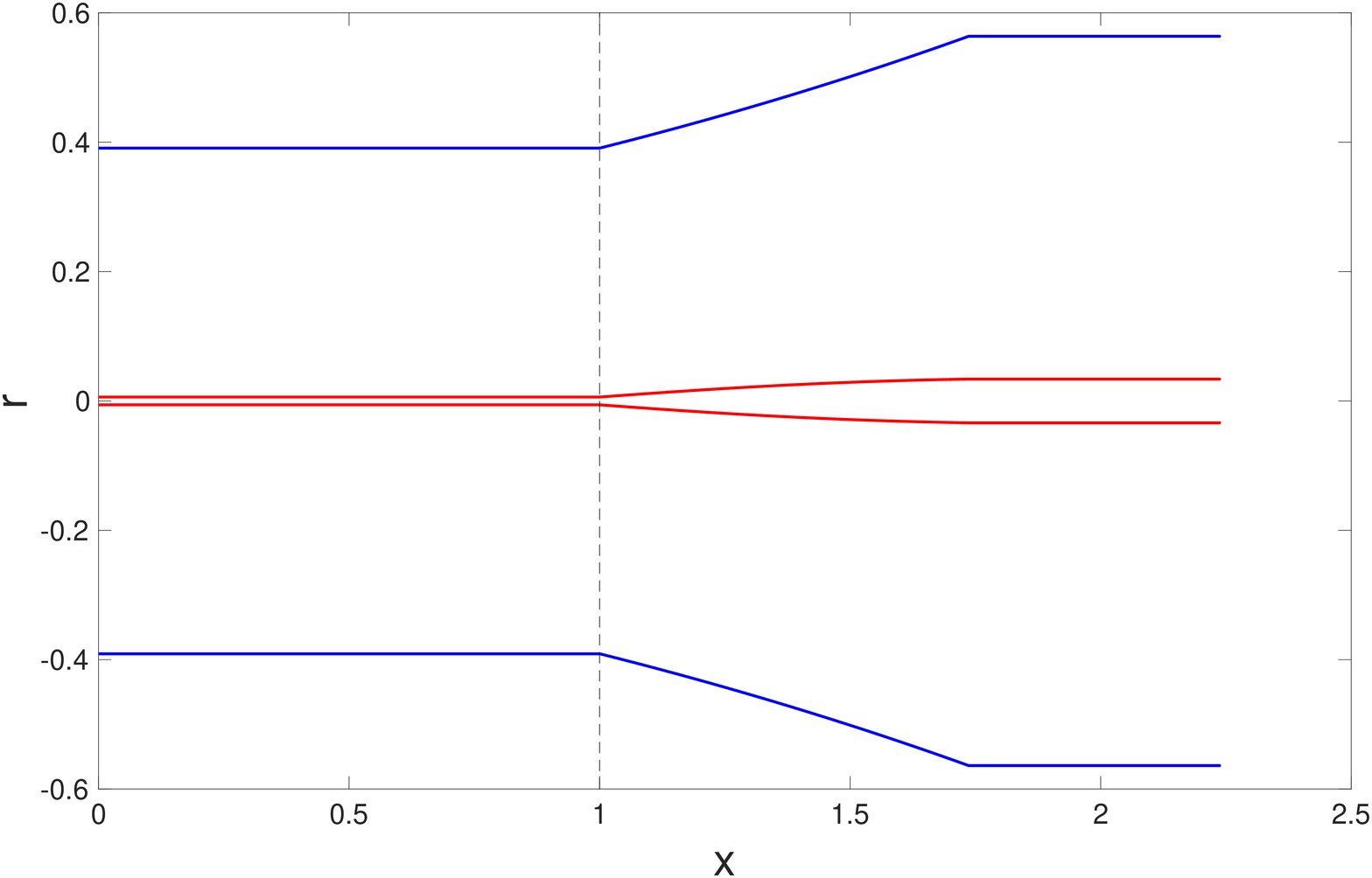}
    \includegraphics[width=0.480\textwidth]{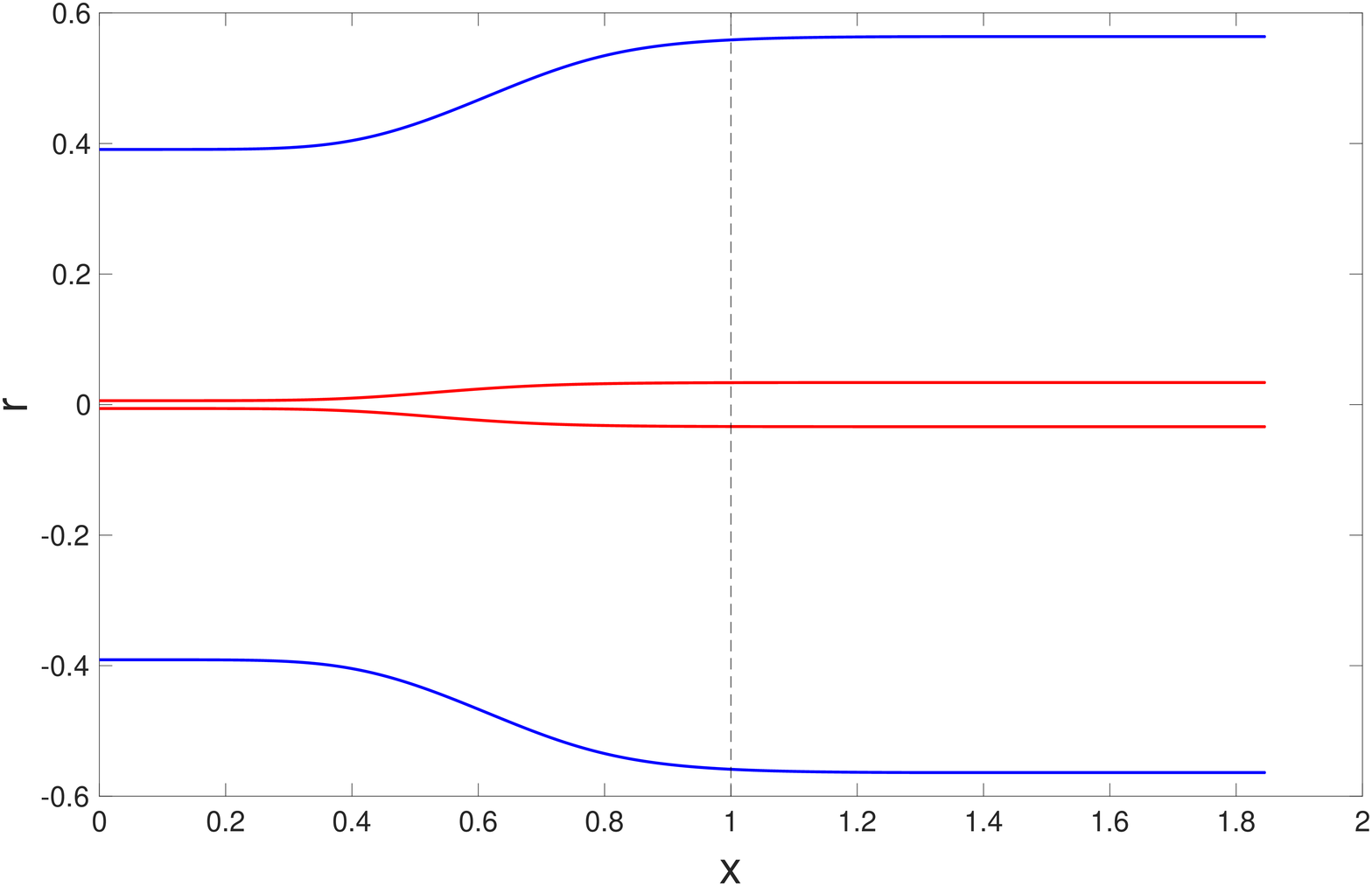}
    \caption{{Emitter tips obtained from an axisymmetric tube with initial length $L(0) = 1.5$, cross-sectional area $\chi^2(x,0)=1$ and aspect ratio $\phi(x,0)=0.06$, by pulling with force $F=2$ and pressure $P=32$ for time $t_{K} = 0.15$. At left is the geometry obtained with the viscosity profile \eqref{viscosity_profile}
    and at right that obtained with the viscosity profile \eqref{realistic_viscosity}. 
    The outer boundary $r=R(x,t_K)$ is shown in blue and the inner channel $r=\phi(x,t_K)R(x,t_K)$ is shown in red.}}
    \label{fig:4}
    \end{figure} 
    
Viscosity plays a significant role in the deformation process, consequently, we will now assume a viscosity profile more likely to correspond to physical reality. The viscosity is expected to vary axially over the heated region, attaining its minimum dimensionless value $\mu=1$ at the centre of the heated region, $x=0$, where the temperature is hottest, and increasing with distance from the centre. At the end of the heated region, $x=1$, the glass will have cooled sufficiently such that it is, effectively, solid with a dimensionless viscosity $\mu_{cold}\gg 1$ and we may take this as the viscosity for $x\ge 1$. We also assume the viscosity is independent of time for the pull time $0\le t\le t_K$. Accordingly, we assume an axially varying viscosity given by
\begin{equation}
\mu (x) = \left\{\begin{array}{ll}
1+(\mu_{cold}-1){x}^n,   & 0\le x\le 1\\
\mu_{cold}, & x>1,
\end{array}
\right.\label{realistic_viscosity}
\end{equation}
where $n$ is a positive integer which permits modification of the rate of increase of the viscosity with $x$; the larger the value, the more uniform the viscosity in the vicinity of $x=0$ and  the more rapid the rise near $x = 1$ which, in turn, leads to a longer taper. Here we take $n=6$. As in 
we take $\mu_{cold} =100$. 

Figure~\ref{fig:4} compares the geometry of emitter tips pulled for a time $t_{K} = 0.15$ with force $F=2$ and pressure $P=32$, and with the viscosity given by the piecewise constant function 
and the more realistic continuous function 
The piecewise constant viscosity results in a longer taper with a uniform tubular geometry over the full heated length $0\le x\le 1$.

\subsection{Achieving a constant bore}

\begin{figure}
    \centering
    \includegraphics[width=0.80\textwidth]{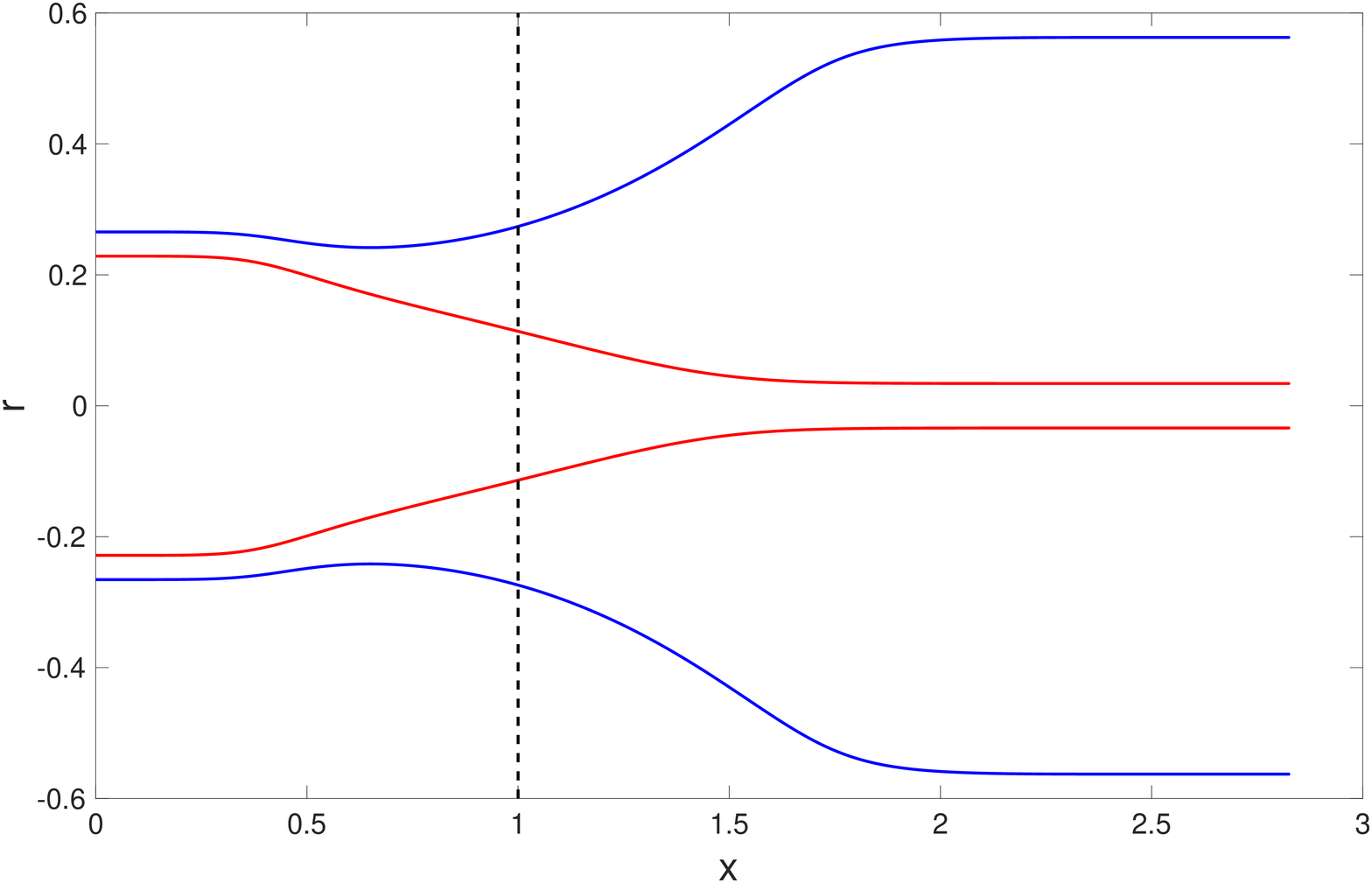}
    \caption{The outer boundary $r=R(x,t)$ (blue) and inner boundary $r=\phi(x,t) R(x,t) $ (red) over the position after pulling from an initial length of $L(O) = 1.5$, cross-sectional area $\chi^2(x,0)=1$, and aspect ratio $\phi(x,0)=0.06$ for an internal channel pressure $P=34.217$, force $F=1.32$, and a draw time $t_{K} = 0.44$. }
    \label{fig:6}
    \end{figure}

As seen in figure~\ref{fig:4}, for the chosen parameters the internal channel radius reduces as $x\rightarrow 0$. This may be reversed by increasing the pressure. Figure~\ref{fig:6} shows the geometry obtained with $P=34.217$, a pulling time of $t_K=0.44$, and with the viscosity given by \eqref{realistic_viscosity}. For these parameters the internal channel radius increases as $x\rightarrow 0$. For a pressure only a little larger, the tube will burst at $x=0$, i.e. the aspect ratio will increase to $\phi=1$. As discussed earlier, it is desirable to produce emitter tips for mass spectrometry with a (nearly) constant bore and with a sharp tip such that the wall thickness at $x=0$ is small. 

We denote $\phi_T=\phi(0,t_K)$ and $\chi_T^2=\chi^2(0,t_K)$ as the aspect ratio and cross-sectional area at position $x=\tilde x=0$ and time $t_K$, from which the external radius $R_T=R(0,t_K)$ is readily determined. Then, assuming a known viscosity $\mu(x)$, for the desired geometry at $x=0$, we must find a force $F$, pressure $P$ and pulling time $t_K=t(0,\tau_K)$ that gives large $\phi_T$ sufficiently close to, but smaller than, unity and small $\chi_T$ sufficiently close to, but larger than, zero. This means we are seeking parameters near to the limiting cases of bursting and breaking where the solution is highly sensitive to small changes in the parameter values making this a challenging problem. Moreover there is not necessarily a unique solution.

\begin{figure}
\centering
    \includegraphics[width=0.80\textwidth]{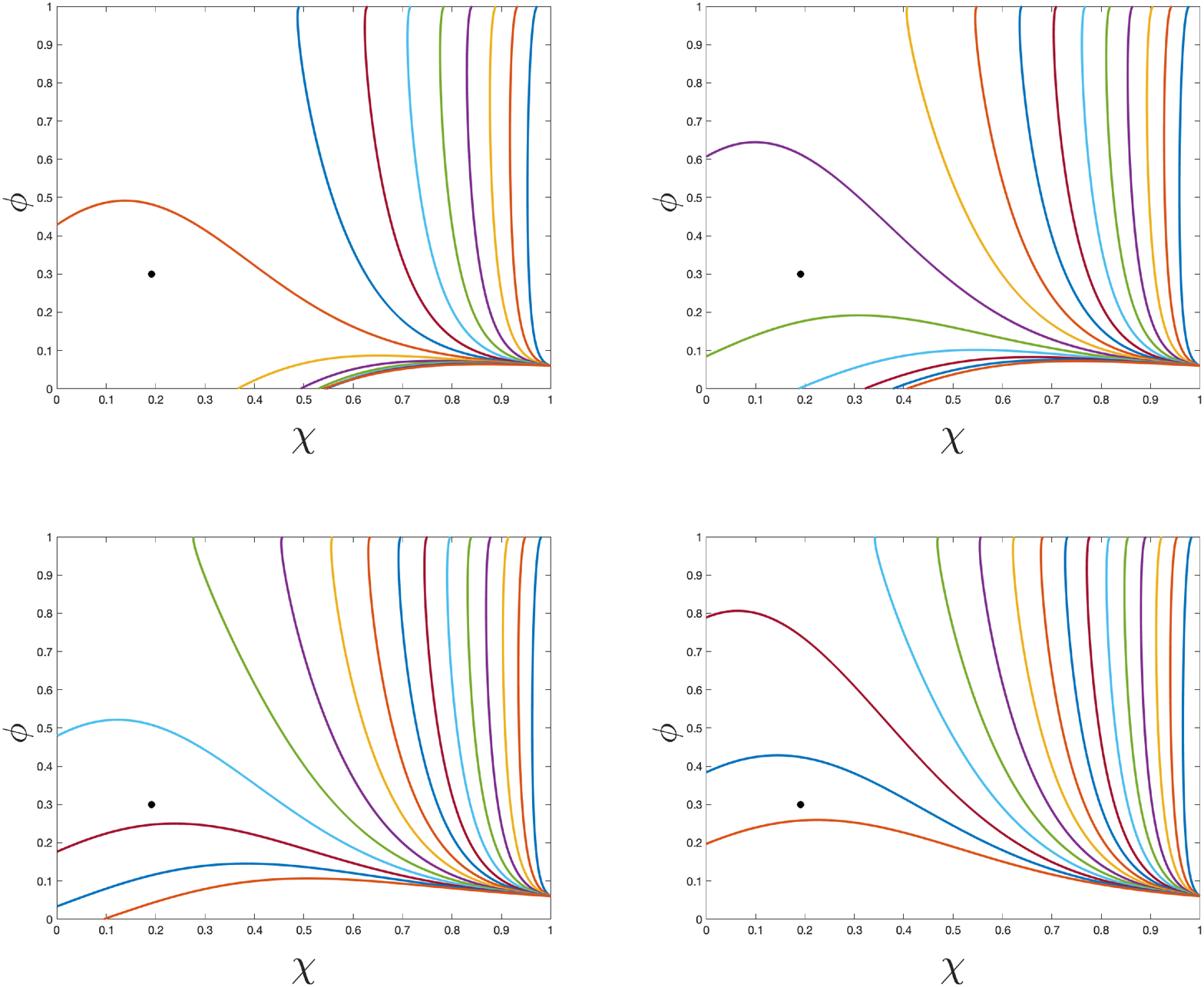}
    \caption{$\phi$ versus $\chi$ for a capillary with initial aspect ratio $\phi_0=0.06$.  Pulling with pressure $P=34$ (top left), $35$ (top right), $36$ (bottom left), $37$ (bottom right) for a force $F$ ranging from $0.5$ to $2$ in increments of $0.1$. The black dot is $(\chi_{T},\phi_T)=(0.1911,0.3)$ corresponding to the the final emitter geometry at $x=0$.}
     \label{fig:7}
    \end{figure}

To aid understanding, we show a number of phase plane plots in figure~\ref{fig:7} for a tube with initial aspect ratio $\phi_0=0.06$, hence (dimensionless) external radius $R_0=0.5652$ (unit cross-sectional area). Each plot is for a different value of pressure $P$. Each of the curves in a plot corresponds to a different pulling force $F$ and shows how the geometry  $(\chi,\phi)$ of every cross-section changes as $\tau$ (and hence $t$) increases, where $(\chi,\phi)=(1,\phi_0)$ at $t=\tau=0$. Each curve stops at one of the boundaries $\phi=0$ (hole closure), $\chi=0$ (breaking of the tube), $\phi=1$ (bursting of the tube), or $\chi=1$ (cross-sectional area unchanged from initial value). These plots show that there are multiple choices of $P$, $F$ and $\tau_K$ that will yield a desired final geometry $(\chi_T,\phi_T)$, where $(\chi_T,\phi_T)=(0.1911,0.3)$ is the black dot in the plots which corresponds to $R_T=0.1130$, $\phi_TR_T=\phi_0R_0=0.0339$, i.e. the internal radius at $x=0$ equals that of the initial tube. 

\begin{figure}
  \centering
  \includegraphics[width=0.80\linewidth]{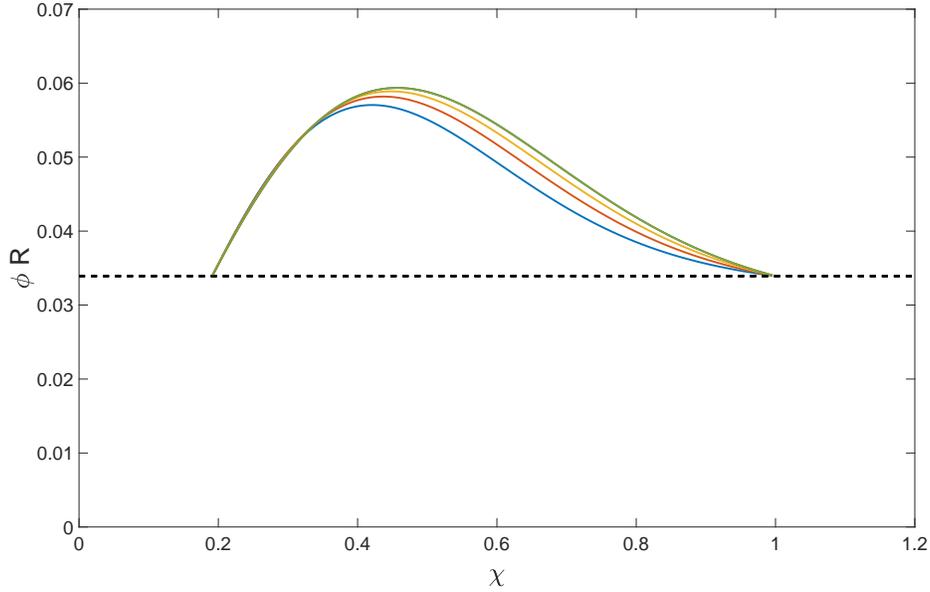}
  \caption{Plot of the internal boundary $r=\phi R$ as a function of $\chi\in[\chi_T,1]$ for (bottom to top) $P=34,\,35,\,36,\,37$ and corresponding $F=1.3166,\,1.5597,\,1.7730,\,1.9690,$  respectively (to  four decimal places), 
  for a tube with initial aspect ratio $\phi_{0}=0.06$. 
  The black dashed line shows the initial inner radius (when $\chi=1$) which is also the radius when $\chi=\chi_T=0.1911$.}
\label{fig:8}
\end{figure}

\begin{figure}
  \centering
  \includegraphics[width=0.80\linewidth]{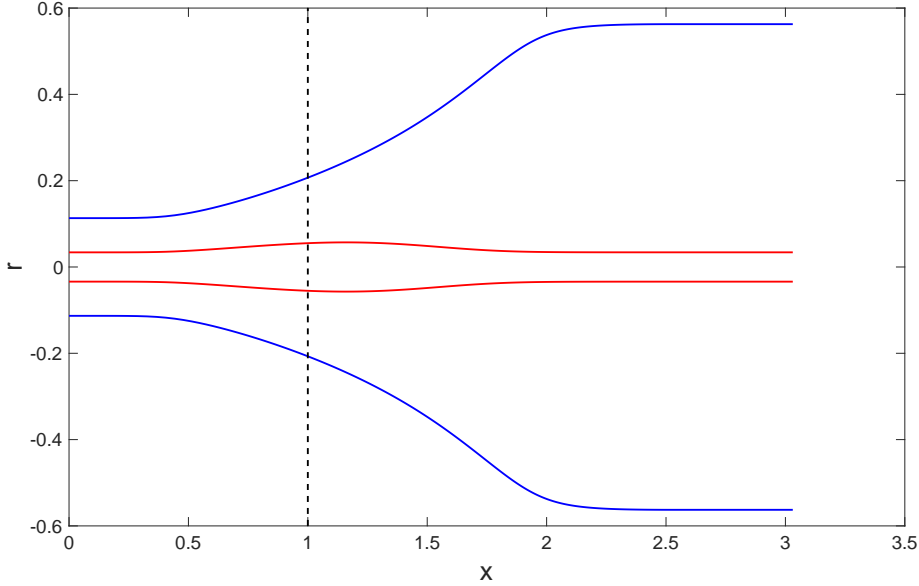}
  \caption{Plot of the final emitter geometry for a tube with initial length $L(0) = 1.5$, cross-sectional area $\chi^2(x,0)=1$, and aspect ratio $\phi(x,0)=0.06$, after pulling for a time $t_{K} = 0.4436$ with force $F=1.3166$ while applying an internal channel pressure $P=34$. The external boundary $r=R(x)$ is shown in blue and the internal boundary $r=R(x)\phi(x)$ is shown in red.
  The aspect ratio and cross-sectional area at the tip are $\phi_{T} =0.3$, $\chi_{T} = 0.1914$. 
  }
\label{fig:9}
\end{figure}

Although there are many combinations of pressure and force 
able to yield the same geometry at $x=0$,
the evolution of the geometry is unique from one solution to another which affects the final geometry of the cross-sections for $x>0$ and, therefore, the shape of the emitter. Therefore, because we are seeking an emitter with constant bore, after determining values $P$ and $F$ corresponding to the desired $\chi_{T}$ and $\phi_{T}$, we look for the best choice of these parameters corresponding to the most uniform bore. Figure~\ref{fig:8}, shows the internal radius $\phi R$ against $\chi\in[\chi_T,1]$ for a number of different parameter combinations $P$, $F$, which shows the change of the internal radius as the cross-sectional area reduces from $\chi^2=1$. Every cross-section will follow the curve corresponding to the chosen parameters $P$, $F$, though only that at $x=0$ will reach the end point $\chi=\chi_T$.
The best choice is the parameter combination that gives the least variation in the inner channel radius. 
Of the curves shown, the best is clearly for $(P,F)=(34,1.3166)$. Figure~\ref{fig:9} shows the shape of the glass emitter for this choice (recall that the axial length scale is $\sim 3\,\text{mm}$ while the radial length scale is $0.1\,\text{mm}$). 

\begin{figure}
  \centering
  \includegraphics[width=0.80\linewidth]{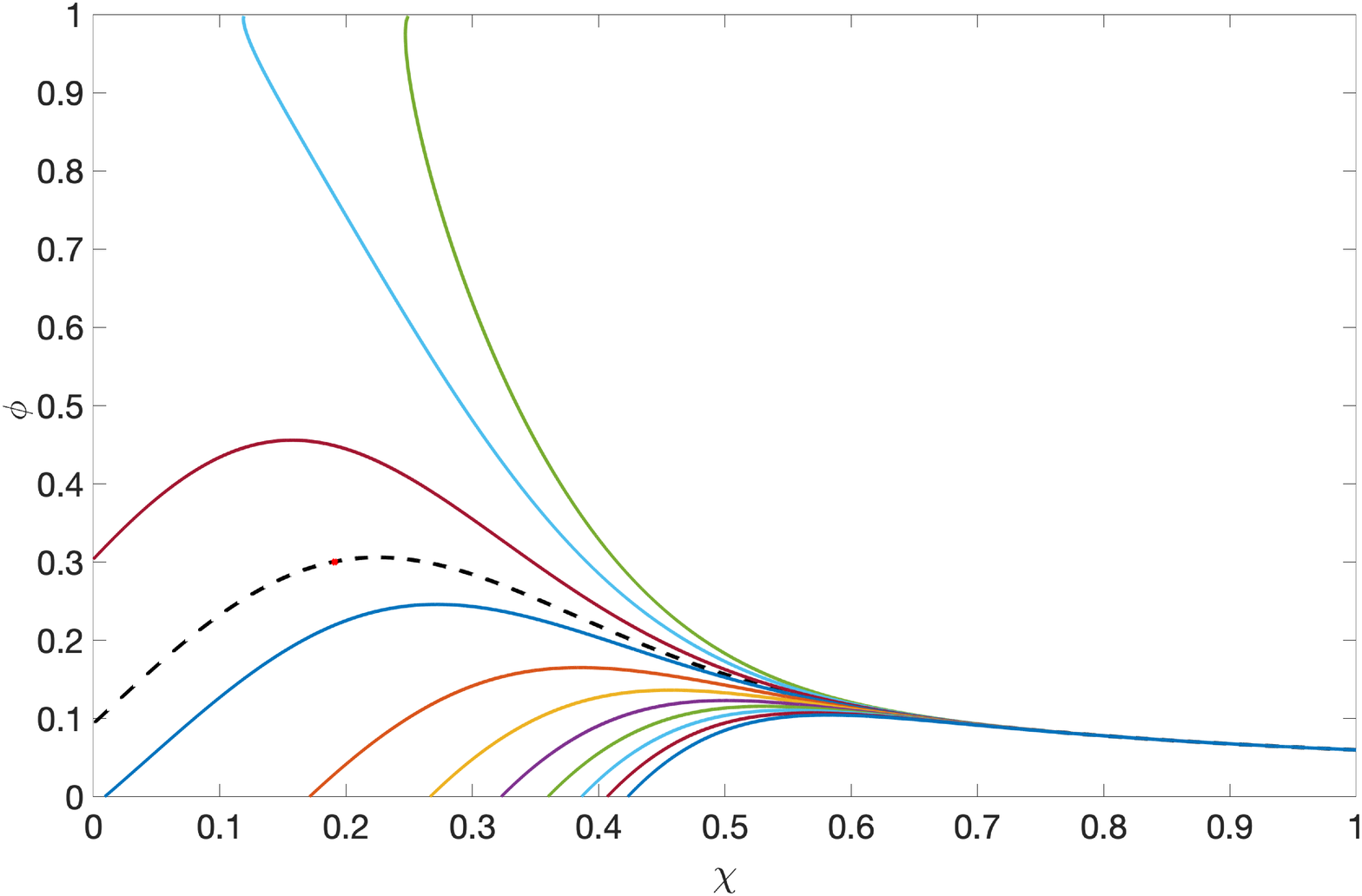}
  \caption{Curves $(\chi(\tau), \phi(\tau))$ for $P=32$ and (solid curves) $0.64102 \le F \le 0.64103$ in increments of $10^{-6}$, with $F$ increasing from the top curve to the bottom curve. Initially $(\chi(0),\phi(0))=(1,0.06)$ and $\tau$ increases from right to left. The red dot shows the desired geometry $(\chi_{T},\phi_{T}) \approx (0.1911,0.3)$ at the tip $x=0$. The black-dashed curve passing through the red dot corresponds to a force of $F = 0.64102262$.}
\label{fig:11}
\end{figure}

\begin{figure}
  \centering
  \includegraphics[width=0.80\linewidth]{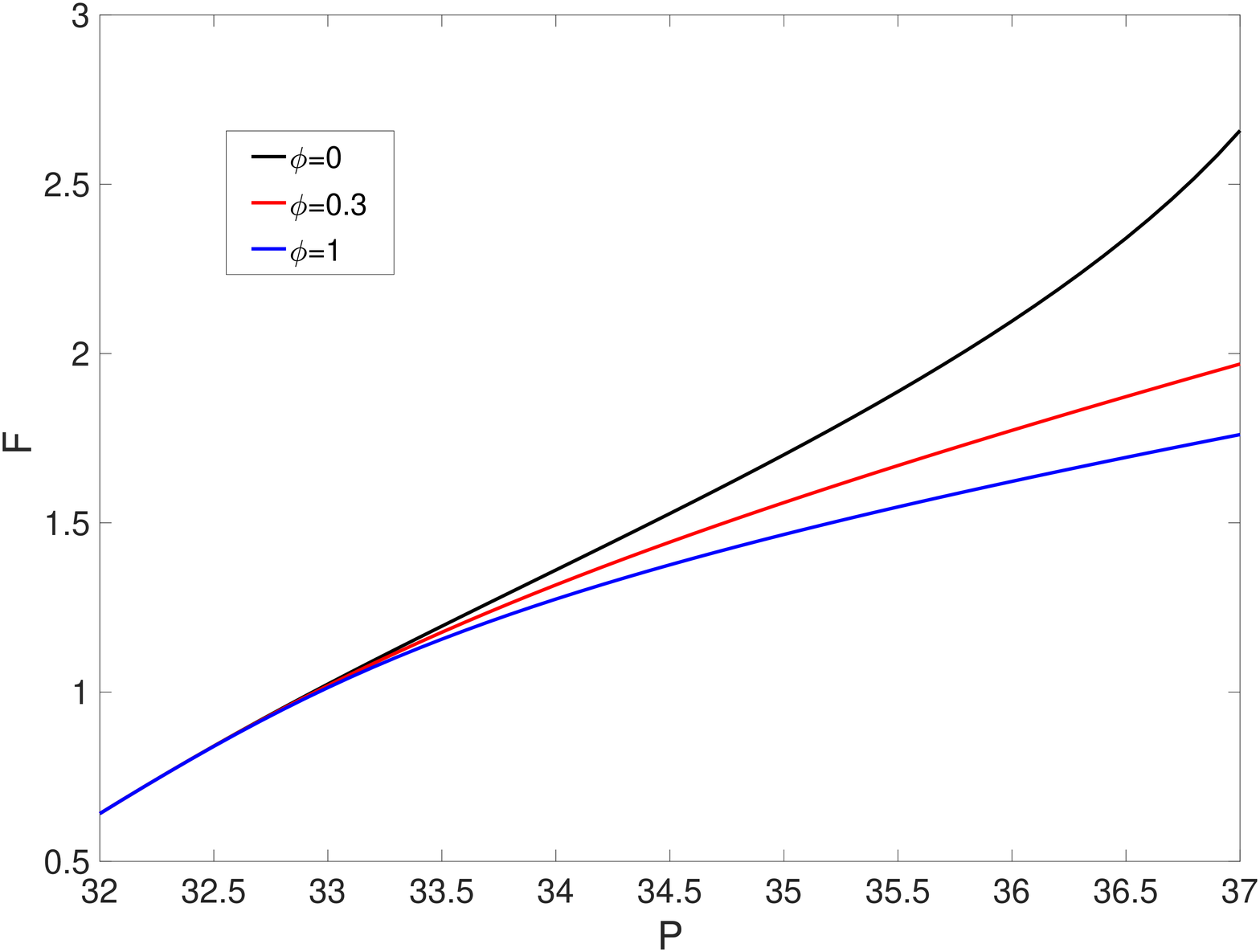}
   \caption{Pulling force $F$ versus pressure $P$ for the manufacture of emitters with $\chi_T=0.1911$ and 
  $\phi_{T} = 0$ (black), $\phi_{T} = 0.3$ (red) and $\phi_{T} = 1$ (blue), from a tube with initial geometry $(\chi,\phi)=(1,0.06)$. }
\label{fig:10}
\end{figure}

Although figure~\ref{fig:8} suggests that the variation in the inner channel radius decreases with decreasing pressure (and pulling force), there is a lower limit 
for a practically feasible outcome. Consider figure \ref{fig:11} showing curves $(\chi,\phi)$ for pressure $P=32$ and different values of force $F$ to which $\phi$ is highly sensitive. To achieve the desired geometry ($\chi_T,\phi_T$) at the tip (red dot through which the black dashed curve passes) a force $F=0.64102262$ is required; for the same $\chi_T$, a small change in the force results in a significant change in $\phi_T$, with the tube bursting ($\phi=1$) for a slightly smaller force in the range $0.641020<F<0.641021$, and the channel closing ($\phi=0$) for a slightly larger force in the range $0.641023<F<0.641024$. Thus, if the necessary (seven figure) accuracy in the force cannot be practically achieved, a geometry vastly different from that desired is highly likely. This implies the need to work in a parameter regime where there is less sensitivity to small changes in the parameters such that the desired geometry at the tip can be assured within reasonable tolerances.  

Figure \ref{fig:10} shows, in red, the curve $(P,F)$ corresponding to $(\chi_{T},\phi_{T}) = (0.1911,0.3)$ from a tube with initial geometry $(\chi,\phi)=(1,0.06)$. Above this is shown, in black, the curve on which $(\chi_{T},\phi_{T}) = (0.1911,0)$, i.e. the channel closes at $x=0$, while below is shown, in blue, the curve on which $(\chi_{T},\phi_{T}) = (0.1911,1)$, i.e. the tube bursts at $x=0$. To obtain a valid emitter tip, even if not the desired geometry, values of $P$ and $F$ must be chosen from the region between the upper and lower curves. 
This figure clearly shows the sensitivity of $\phi_T$ to the force $F$ for smaller pressure $P$, and the decrease in sensitivity as $P$ increases. However, while choosing larger pressure and force means less sensitivity of the tip geometry to inaccuracies in the parameters, it needs to be remembered that the variation of the inner channel radius increases, and a larger bulge is seen; see figure \ref{fig:8}. We conclude that the optimal choice of the parameters $P$, $F$ corresponds to the point on the red curve with smallest $P$ such that variations within the tolerance range of the pulling device lead to acceptable variation in the geometry at $x=0$. This will yield the emitter with the most uniform internal channel radius (smallest bulge) possible.

\section{Conclusion}\label{sec:conclusions}

We have developed a mathematical model and numerical solution method of the unsteady heat and draw process used for tapering of tubular glass fibres. For the first time active pressurisation of internal channels has been included in an unsteady fibre-drawing  model, enabling investigation of the competition between surface tension and pressure. 
We have considered the  effects on the final geometry of pulling force, surface tension, active channel pressurisation, and axially varying viscosity along the heated region. Our primary focus has been the manufacture of tapers for use in mass spectrometry having a near-uniform bore and a small wall thickness at the very tip. The use of an over-pressure in the channel, to counteract the reduction in its size as the cross-sectional area decreases due to pulling and the channel closes due to surface tension, has been of particular interest. 

The most important outcome of this work is to demonstrate the feasibility of manufacture of tapers with near-uniform bore by the heat and pull process with active pressurisation of the channel. The symmetry of the process means that two identical tapers are obtained from a single heat and pull process by cutting/breaking the tube at the centre. The geometry at the centre then corresponds to the geometry at the very tip of the taper. The model and solution method here described, enable determination of a pulling force $F$, channel over-pressure $P$, and draw time $t_K$, to achieve tapers with a desired internal diameter and wall thickness at the very tip from a given tubular fibre for a prescribed viscosity profile that may vary in time and space. There is not a unique solution to this problem; multiple parameter sets  $(F,P,t_K)$ will yield the same geometry at the tip with a larger pulling force requiring a larger over-pressure. However, the geometry along the length of the taper, which is also found, will vary between parameter sets and this may be used to determine the best choice of the parameters. 

The geometry of the taper is quite sensitive to the pulling force and pressure parameters. A taper for use in mass spectrometry, having a tip with channel diameter near to that of the original fibre and small wall thickness, implies a geometry at the tip with significantly larger aspect ratio $\phi$ and smaller cross-sectional area $\chi^2$ than the original fibre which, in turn, implies an over-pressure and  pulling force that are nearing the regime where the fibre may burst or break. As the pulling force and over-pressure reduce, the sensitivity increases such that a very small variation in the force or pressure will result in the fibre bursting ($\phi\rightarrow 1$) or the channel closing ($\phi\rightarrow 0$). Note that such a failure will occur first at the tip which is subjected to the greatest amount of heat and consequently undergoes more deformation than any other cross-section.  Practical limitations on maintaining the force and pressure at given values will, then, require that these parameters be sufficiently large so that small variations do not result in unacceptable changes in $\phi$ at the tip. However, larger force and over-pressure result in greater non-uniformity of the internal channel, i.e. a larger bulge in the inner channel, so that increasing these parameters too much is also not desirable. Clearly there is a trade-off between the two and an optimisation problem to be solved based on physical limitations of the puller and fitness for purpose of the taper.

In this paper we have assumed an axial viscosity profile which is highly dependent on temperature which is difficult to measure. Thus, the flow model here described needs to be coupled with an energy model in a similar manner to \citet{stokes2019coupled}. However, the energy model will depend on the puller used and we have left this to a future publication. While the temperature profile will impact the geometry of the taper, it will not change the fundamental findings of this work, namely that including over-pressure in the unsteady heat and pull process provides an important additional control for obtaining a desired geometry and that physical tolerances on the draw parameters and acceptable tolerances on the final geometry need to be considered to find the optimal choice of draw parameters.

We conclude by noting that, while we have focused on manufacture of emitters for mass spectrometry, our model has wider application such as to the manufacture of microbottles for whispering gallery resonator sensors.

\begin{acknowledgements}
This work was supported by the Australian Research Council grant FT160100108.
\end{acknowledgements}

\bibliographystyle{plainnat}
\bibliography{Paper_v1}

\end{document}